\documentstyle[12pt,aps]{revtex}
\begin{document}
\title{
\hfill\parbox[t]{2in}{\rm\small\baselineskip 14pt
{JLAB-THY-98-48}}
\bigskip\bigskip
\vskip 2cm
Interpreting the Neutron's Electric Form Factor:\\  
Rest Frame Charge 
Distribution or Foldy Term?}

\vspace{2.0 cm}
\author{Nathan Isgur}
\address{Jefferson Lab\\ 12000 Jefferson Avenue,
Newport News, Virginia, 23606}
\maketitle

\vspace{2.0 cm}
%%%%%%%%%%%%%%%%%%%%%%%%%%%%%%%%%%%%%%%%%%%%%%%%%%%%%%%%%%%%%%%%%%%%
\begin{center}  {\bf Abstract}  \end{center}
%%%%%%%%%%%%%%%%%%%%%%%%%%%%%%%%%%%%%%%%%%%%%%%%%%%%%%%%%%%%%%%%%%%%
\vspace{.4 cm}
\begin{abstract}

		The neutron's electric form factor contains vital information
on nucleon structure, but its interpretation within many models
has been obscured by relativistic effects. I demonstrate  that,
to leading order in the relativistic expansion of a constituent quark model, 
the Foldy term cancels exactly
against a contribution to the Dirac form factor $F_1$ to leave intact the naive interpretation
of $G^n_E$ as arising from the neutron's rest frame charge distribution.
\bigskip\bigskip\bigskip\bigskip

\end{abstract}
\pacs{}
\newpage

%%%%%%%%%%%%%%%%%%%%%%%%%%%%%%%%%%%%%%%%%%%%%%%%%%%%%%%%%%%%%%%%%%
\section {Introduction}
%%%%%%%%%%%%%%%%%%%%%%%%%%%%%%%%%%%%%%%%%%%%%%%%%%%%%%%%%%%%%%%%%%
\medskip
	In 1962, Sachs showed $\cite{Sachs}$ that the combinations of elastic nucleon
form factors ($N=p$ or $n$)
\begin{equation}
G^N_E = F^N_1 - {Q^2 \over 4m_N^2}F_2^N
\end{equation}
\begin{equation}
G^N_M = F^N_1 + F_2^N
\end{equation}
have simple interpretations as the spatial Fourier transforms of the nucleons' charge and
magnetization distributions in the Breit frame (where momentum $\vec p= - {\vec Q \over 2}$
is scattered to momentum $\vec p^{~\prime}= +{\vec Q \over 2}$).  Here $F^N_1$ and $F^N_2$ are the
Dirac and Pauli form factors, respectively, defined by 
\begin{equation}
\langle N(\vec p^{~\prime},s') \vert j^{\mu}_{em}(0) \vert N(\vec p,s) \rangle =
\bar u(\vec p^{~\prime},s')[F^N_1 \gamma^{\mu} + i{\sigma^{\mu \nu}q_{\nu} \over 2m_N}F^N_2]u(\vec p,s)
\end{equation}
where $q_{\nu}=p^{~\prime}_{\nu}- p_{\nu}$ and $F^N_1$ and $F^N_2$   are functions of $Q^2 = -q^2$.

	These
form factors obviously contain vital information on the internal composition
of the nucleons.  Although it has proven elusive experimentally, the electric
form factor of the neutron $G^n_E$ is particularly fascinating in this respect.
In pion-nucleon theory, $G^n_E$ would arise from a $\pi^-$ cloud with convection
currents producing the anomalous magnetic moments 
$F^p_2=1.79\equiv \mu_p-1$ and $F^n_2=-1.91\equiv \mu_n$.
In contrast, in a valence quark model, the nucleon magnetic moments arise from the underlying charged
spin-$1\over 2$ constituents with the famous $SU(6)$ relation
\begin{equation}
{\mu_p \over \mu_n}=-~{3 \over 2}
\end{equation}
and with a scale set by
\begin{equation}
\mu_p={m_N \over m_d}\simeq 3
\end{equation}
where $m_d \simeq m_u \simeq {1 \over 3}m_N$
is a valence quark effective mass.  Within this model it has been argued that in the $SU(6)$ limit  
$G^n_E (Q^2)$ would be identically zero, 
but that the spin-spin forces which produce the $SU(6)$-breaking $\Delta-N$ splitting create a charge 
segregation inside the neutron and lead to a nonzero 
$G^n_E$ \cite{qmneutron1,qmneutron2,qmneutron3}.  The effect arises because the spin-spin 
forces push $d$ quarks to the periphery of the neutron and pull the $u$ quark to the 
center.  Thus both the $\pi^-$ cloud picture and the hyperfine-perturbed quark model
predict a negative neutron charge radius, as observed.

		Nonrelativistically, the squared charge radius is simply the charge-weighted
mean square position of the constituents.  More generally
\begin{equation}
G^p_E(Q^2) \equiv 1-{1 \over 6} r^2_{Ep}Q^2+ \cdot \cdot \cdot
\end{equation}
and
\begin{equation}
G^n_E(Q^2) \equiv -{1 \over 6} r^2_{En}Q^2+ \cdot \cdot \cdot
\end{equation}
define the proton and neutron charge radii,
with
\begin{equation}
{G^N_M(Q^2) \over \mu_N}=1-{1 \over 6} r^2_{MN}Q^2 + \cdot \cdot \cdot
\end{equation}
defining the corresponding magnetic radii.  Experimentally, the three form factors
$G^p_E$, $G^p_M$, and $G^n_M$ are reasonably well known, although new
measurements should soon determine them with much greater precision \cite{Perdrisat}.  From
these measurements, which cover $Q^2$ in the multi-GeV$^2$ range, we
have learned that all three form factors have roughly similar shapes:
\begin{equation}
{G^p_E(Q^2)}
\simeq {G^p_M(Q^2) \over \mu_p}
\simeq {G^n_M(Q^2) \over \mu_n}
\simeq G_D(Q^2)
\end{equation}
where the dipole form factor
\begin{equation}
G_D(Q^2) \equiv {1 \over [1+Q^2/M_{dipole}^2]^2}
\end{equation}
with $M_{dipole}^2=0.71$ GeV$^2$.  I note in passing that
this observation makes little sense in a picture where the nucleons have a point-like
core surrounded by a pion cloud (since in such a  picture $G^n_M$ is purely pionic
while $G^p_E$ and $G^p_M$ are mixtures of a ``bare" proton and a pion cloud).  In contrast,
Eq. (9) is very natural in a valence quark model where, 
to leading order in a relativistic expansion, $G_D(Q^2)$ would simply be the
Fourier transform of the square of the ground state spatial wavefunction.

	Recent advances in experimental technique should lead to a clear measurement
of the neutron's electric form factor $G^n_E(Q^2)$ for $Q^2$ in the GeV$^2$ range in the next
few years \cite{GEnreview}.  Its charge radius $r^2_{En}$ 
is known from low energy neutron-electron elastic
scattering to be $-0.113 \pm 0.005$ fm$^2$ \cite{nescattering}, but the electron-neutron 
scattering measurements needed to determine $G^n_E$ in the GeV$^2$ range
(in order to roughly map out the neutron's electric structure with a resolution 
of 10\% of the proton's size) have been plagued by the lack of a free neutron
target \cite{GEnreview}.    
Fortunately, the method of Arnold, Carlson, and Gross \cite{ACG}, which
uses spin observables sensitive to $G^n_E - G^n_M$ interference, 
opens up new methods for measuring $G^n_E$, and recent advances
in accelerator, target, and detector technology are beginning
to exploit these new methods.  

In anticipation of these measurements,
there has been renewed discussion about their interpretation.  I focus here on the belief that the measured
$r^2_{En}$ is explained by the ``Foldy term" \cite {Foldyconfusion}.  {\it I.e.}, using Eqs. (1) and (7), 
\begin{equation}
r^2_{En}=r^2_{1n}+{3 \mu_n \over 2 m^2_N} \equiv r^2_{1n} + r^2_{Foldy,n} ~~~,
\end{equation}
where $r^2_{1n}$ is the ``charge radius" associated with 
$F^n_1 \simeq - {1 \over 6} r^2_{1n}Q^2 + \cdot \cdot \cdot$.  The second term  in Eq. (11), 
called the Foldy term,
appears to arise as a relativistic correction associated with the neutron's magnetic moment and so to have nothing to
do with the neutron's rest frame charge distribution. It has the value $-0.126$ fm$^2$, nearly
coinciding with the measured value. 
On this basis it has been argued that any ``true" charge distribution effect must be
very small.  In this paper I will show that while the Foldy term closely resembles
$r^2_{En}$ numerically, it does not ``explain it".  Indeed, I will
demonstrate that, in the relativistic approximation to the constituent quark model in which the Foldy term first
appears, it is cancelled exactly by a contribution to the Dirac form factor $F_1$
leaving $r^2_{En}$  correctly interpreted as arising entirely from the rest frame internal charge
distribution of the neutron. 
\bigskip
\bigskip

%%%%%%%%%%%%%%%%%%%%%%%%%%%%%%%%%%%%%%%%%%%%%%%%%%%%%%%%%%%%%%%%%%%%%%%%%
\section {The Interpretation of the Neutron Charge Radius in a Constituent Quark Model}
%%%%%%%%%%%%%%%%%%%%%%%%%%%%%%%%%%%%%%%%%%%%%%%%%%%%%%%%%%%%%%%%%%%%%%%%%
\medskip
		The relationship (1) between the Sachs form factor $G_E$  and the
Dirac and Pauli form factors $F_1$ and $F_2$ is relativistic
in origin.  Unfortunately,
relativistic constituent models  of the nucleon are notoriously
difficult:  rest frame models are difficult to boost and infinite-momentum-frame
(or light-cone) quark models have trouble  constructing states of definite $J^P$.
This could be the reason that the interpretation of $G^n_E$ has not been clarified
in the context of such models.

     	While an {\it accurate} constituent quark model of nucleon structure must certainly
be fully relativistic, the issue at hand can be resolved by using
a relativistic expansion around the nonrelativistic limit. This is possible
because the Foldy term $r^2_{Foldy,n}$ arises at order $Q^2/m^2$ and so its character
may be exposed by an expansion of $G_E^n$ to order $1/m^2$. I will also
exploit symmetries of the problem available in certain limits which will make
the discussion independent of the details of models.

    I begin with a simple ``toy model" in which a ``toy neutron" $\tilde n_{\bar S D}$
is composed of a {\it scalar} antiquark $\bar S$ of mass $m_S$ and charge $-e_D$ and a
spin-$1\over 2$ Dirac particle $D$ of mass $m_D$ and charge $e_D$ bound 
by flavor and momentum independent forces into a rest frame nonrelativistic $S$-wave.
	The calculation begins by noting that, from their definitions,
\begin{equation}
G_E^{\tilde n_{\bar S D}}(Q^2) = \langle \tilde n_{\bar S D}(+{Q \hat z \over 2},+) \vert
\rho_{em}
\vert \tilde n_{\bar S D}(-{Q \hat z \over 2},+) \rangle
\end{equation}
and
\begin{equation}
G_M^{\tilde n_{\bar S D}}(Q^2) = {m_{\tilde n_{\bar S D}} \over Q}\langle \tilde n_{\bar S D}(+{Q \hat z \over 2},+) \vert
j^{1+i2}_{em}
\vert \tilde n_{\bar S D}(-{Q \hat z \over 2},-) \rangle ~~~.
\end{equation}
It is immediately clear that the calculation of these form factors 
requires boosting the rest frame  $S$-wave bound state to momenta $\pm {Q \hat z \over 2}$.
Doing so can introduce a host of $1/m^2$ effects in the
boosted counterpart of the $S$-wave state and it can also produce new $P$-wave-like components
by Wigner-rotation of the $D$-quark spinors \cite{Wignerrotation}.   
I will show that the latter effect is subleading, and will
deal 
with the former effect by exploiting an effective charge-conjugation symmetry of
the system for $m_D=m_S \equiv m$. 

   Since $\mu_{\tilde n_{\bar S D}}$ involves the limit of Eq. (13) as $Q \rightarrow 0$, 
to the required order in $1/m$ it simply takes on its nonrelativistic value
\begin{equation}
\mu_{\tilde n_{\bar S D}}={e_D m_{\tilde n_{\bar S D}} \over m_D}
\end{equation}
where of course $m_{\tilde n_{\bar S D}}=m_S+m_D$ in this limit. The Foldy term
is thus well-defined:
\begin{equation}
r^2_{Foldy,\tilde n_{\bar S D}}={e_D  \over 2  m_D m_{\tilde n_{\bar S D}}}~~~.
\end{equation}
We next compute $G_E^{\tilde n_{\bar S D}}(Q^2)$ directly from Eq. (12). To 
leading order in $1/m^2$, $\rho_{em}$ remains a 
one-body current, and the impulse approximation is valid. Within this approximation, we make
use of the relation
\begin{equation}
\langle D(\vec p+  Q \hat z,s') \vert \rho_{em} \vert D(\vec p,s) \rangle =
e_D(1-{Q^2 \over 8 m_D^2})
\langle \tilde D(\vec p+Q \hat z) \vert \rho_{em} \vert \tilde D(\vec p) \rangle \delta_{ss'}
+\rho_{spin-flip}
\end{equation}
where
\begin{equation}
\rho_{spin-flip} \equiv {{e_D Q} \over 4 m_D^2} (p_-\delta_{s'+}\delta_{s-}-p_+\delta_{s'-}\delta_{s+})
\end{equation}
and where $\tilde D$ is a fictitious scalar quark with the mass and charge of $D$. This
expression is easily obtained by making a nonrelativistic expansion of both
the $D$ and $\tilde D$ charge density matrix elements.

   The spin-flip term $\rho_{spin-flip}$ can only contribute {\it via} transitions
to and from the Wigner-rotated components
of the wavefunction. However, the amplitudes of such components are proportional to
$Qk/m_D^2$, where $k$ is an internal momentum. Since $\rho_{spin-flip}$ 
already carries a factor $1/m_D^2$, such effects may be discarded. Note that non-flip Wigner-rotated 
contributions are of the same order and may also be neglected.

   We conclude that $r_{E \tilde n_{\bar SD}}^2$ may be computed by
replacing $D$ by $\tilde D$ provided the additional
contribution $-e_DQ^2/8m_D^2$ is added to $G_E^{\tilde n_{\bar S D}}$.
(More precisely, this factor multiplies the $\tilde D$ contribution
to $G_E^{\tilde n_{\bar S D}}$, but at $Q^2=0$ this is just unity.) I will
denote the associated ``zwitterbewegung" charge radius $3e_D/4m_D^2$ by $r_{D,zwitter}^2$. The
effect of $r_{D,zwitter}^2$ is well-known in a variety of contexts, including atomic physics \cite{atoms}, 
nuclear physics \cite{gross}, hadronic physics \cite{HayneIsgur}, and 
heavy quark physics \cite{onequarter}; at the most elementary and concrete
level it appears as the additional factor of $(1-{Q^2 \over 8m_D^2})^2$ in
the ratio of the Mott cross section to the Rutherford cross section. The problem of computing the remaining
contributions to $G_E^{\tilde n_{\bar S D}}$ to this order from the fictitious $\bar S \tilde D$ scalar-scalar
bound state would in general be highly nontrivial. However, in the limit
$m_D=m_S \equiv m$ these contributions {\it vanish}, since this system
has in this limit a pseudo-charge-conjugation invariance
under $(\bar S, \tilde D) \rightarrow  ( S, \bar {\tilde D})$. However, we note that in this limit
$m_{\tilde n_{\bar SD}}=2m$ and so from Eq. (15) we have
\begin{equation}
r^2_{E\tilde n_{\bar S D}} =r^2_{D,zwitter}={3 e_D  \over 4  m^2} = r^2_{Foldy,\tilde n_{\bar S D}}~~~,
\end{equation}
{\it i.e., in this model the ``scalar charge
distribution" is zero and the Foldy term
would indeed account for the full charge radius of $\tilde n$.} This conclusion is simply interpreted: the two scalar
particles $\bar S$ and $\tilde D$ have perfectly overlapping and cancelling charge distributions,
but the expansion of the $\tilde D$ distribution by $r_{D,zwitter}^2$ creates a
slight excess of $\tilde D$ at large radii. In terms of its experimental 
significance, we have concluded that in an $\bar S D$ model of
the neutron, the observation of an equality of
$r^2_{En}$ and $r^2_{Foldy, n}$ would indeed indicate the absence of an intrinsic ``scalar" charge distribution.

  We shall soon be drawing quite another conclusion for the situation 
in the constituent quark model. However, before leaving the $\bar S D$ model, it is useful 
to consider another limit: the ``hydrogenic limit"  where $m_S \rightarrow \infty$. In this case (see Eq. (15)),
$r^2_{Foldy,\tilde n_{\bar S D}}=0$ but
\begin{equation}
r^2_{E,\tilde n_{\bar S D}} = {3 e_D  \over 4  m_D^2}+e_D r^2_{wf}~~~,
\end{equation}
where $r^2_{wf}$ is the charge radius associated with the bound state $\tilde D$ problem.
It may be that $r^2_{wf}$ contains other $1/m_D^2$ effects, 
but we note that the physics of the  $r_{D,zwitter}^2$ effect previously associated with 
$F_2$ {\it via} $r^2_{Foldy,\tilde n_{\bar S D}}$ now must be asociated with $F_1$ {\it via}
$r^2_{1\tilde n_{\bar S D}}$. The $\bar S D$ toy model thus simultaneously supplies us
with a simple interpretation of the Foldy term and a warning about
associating $F_1$  with the neutron's ``intrinsic charge distribution".

   While the $\bar S D$ toy model has some of the characteristics of
diquark models for the nucleon ($\bar S$ has the quantum numbers and color of a scalar diquark),
our main use for it was to introduce the basic elements of our discussion in a
simple context. Indeed, it is now
relatively trivial to extend
our considerations to the  realistic case of the valence quark model in which the
neutron is in the leading approximation
made of three  mass $m_q$ spin-${1 \over 2}$ quarks $ddu$ 
bound by flavor and momentum independent forces into flavor-independent 
nonrelativistic relative $S$-waves. In this case 
\begin{equation}
\mu_{\tilde n_{ddu}}=-{2m_{\tilde n_{ddu}} \over 3m_q} \simeq -2
\end{equation}
so that
\begin{equation}
r^2_{Foldy,\tilde n_{ddu}}=-{1 \over m_qm_{\tilde n_{ddu}}}~~~.
\end{equation}
In calculating $r^2_{E\tilde n_{ddu}}$   {\it via} Eq. (12), the
transformation of the calculation
of the charge radius of $ddu$ to that of three scalar quarks $\tilde d \tilde d \tilde u$ and residual
$r^2_{q,zwitter}=-e_qQ^2/8m_q^2$ terms proceeds as before, as does the neglect
of Wigner-rotated components of the boosted state vectors.
{\it However, in this case, since the $r^2_{q,zwitter}$ terms are spin and flavor independent, and since
the sum of the three charges is zero, they lead to  no net $Q^2/m^2$ term!} The reason for this is clear: the exactly overlapping
and cancelling quark distributions remain exactly overlapping and cancelling after they are all equally smeared
by $r^2_{q,zwitter}$ \cite{protoncomment}. This picture also anticipates the next stage of the argument: the analog
of the pseudo-charge-conjugation invariance that
we used for the scalar part
of the $\bar S D$ matrix element is that the three quark wavefunction belongs to
the symmetric representation of the permutation
group $S_3$ so that the scalar part $r^2_{E, \tilde n_{\tilde d \tilde d \tilde u}}$
of the charge radius vanishes. Thus in the
usual valence quark
model 
\begin{equation}
r^2_{E,\tilde n_{ddu}}=r^2_{E, \tilde n_{\tilde d \tilde d \tilde u}} + \Sigma_i e_i r^2_{i,zwitter} = 0~~~,
\end{equation}
which requires that
\begin{equation}
r^2_{1 \tilde n_{ddu}} = -r^2_{Foldy, \tilde n_{ddu}}~
\end{equation}
so {\it it is appropriate to interpret the
observed $r^2_{En}$ as due to an intrinsic internal charge distribution}. Stated
in another way, the coincidence of the predicted rest frame charge distribution in such models with 
the experimental value of $r^2_{En}$ may be claimed as a success, while the
numerical coincidence of $r^2_{En}$ with the Foldy term may consistently be viewed as a 
potentially misleading accident. Such an accident is possible because while
in the nonrelativistic limit $r^2_{Foldy,\tilde n} << r^2_{E \tilde n}$, in QCD both constituent masses and hadronic radii
are determined by $\Lambda_{QCD}$ so they are {\it expected} to be of comparable magnitude.

%%%%%%%%%%%%%%%%%%%%%%%%%%%%%%%%%%%%%%%%%%%%%%%%%%%%%%%%%%%%%%%%%%%%%%%%%
\section {Conclusions and Final Remarks}
%%%%%%%%%%%%%%%%%%%%%%%%%%%%%%%%%%%%%%%%%%%%%%%%%%%%%%%%%%%%%%%%%%%%%%%%%
\medskip
	The principal conclusion of this paper is that, within the context of the valence
quark model, the apparent contribution $3\mu_n /2 m_N^2$ of the Foldy term to the charge
radius of the neutron is illusory:  in the leading approximation it is exactly
cancelled by a ``nonintuitive" contribution to the radius $r^2_{1n}$ of the Dirac form
factor $F^n_1$.  It is therefore totally appropriate to compare the measured
$r^2_{En}$ and in general $G^n_E(Q^2)$ against quark model predictions (see, {\it e.g.}, Fig. 1 of 
the second of Refs. \cite{qmneutron2})
for the rest
frame internal charge distribution of the neutron.  

	Before too much is made of this successful prediction of the valence quark model, some other very fundamental
questions must still be answered.  Perhaps the most fundamental
is the possible effect of nonvalence components in the neutron wavefunction.
After all, the classic explanation \cite{picloud} for $r^2_{En}$ is that the neutron has 
a $p \pi^-$ component in its wavefunction \cite{ppicomment} (for a discussion in the more modern context 
of heavy baryon chiral perturbation theory, see Ref. \cite{hbchpt}).  
Since both hyperfine interactions and $q \bar q$ pairs
are $1/N_c$ effects, I know of no simple argument for why one should dominate.

	Fortunately, there is both theoretical and experimental progress in resolving
this old question.  Recent theoretical work on ``unquenching the quark model"
\cite{GI} indicates that there are strong cancellations between the
hadronic components of the $q \bar q$ sea which tend to make it transparent to photons.
These studies provide a natural way of understanding the successes of the valence
quark model even though the $q \bar q$ sea is very strong, and in particular suggest that
the precision of the OZI rule is the result of {\it both} a factor of $1/N_c$ and
strong cancellations within this $1/N_c$-suppressed meson cloud.  New data \cite{HAPPEX} on the
contributions of $s \bar s$ pairs to the charge and magnetization distribution of the 
nucleons is also beginning to constrain the importance of such effects and, by broken $SU(3)$,
their $u \bar u$ and $d \bar d$ counterparts, and future experiments will either see
$s \bar s$ effects or very tightly limit them (at the level of contributions of a  few
{\it percent} to $r^2_E$ and $\mu_N$).  The resolution of the old question
of the origin of $\mu_n$ and $r^2_{En}$ is thus within sight.

\bigskip\bigskip

%%%%%%%%%%%%%%%%%%%%%%%%%%%%%%%%%%%%%%%%%%%%%%%%%%%%%%%%%%%%%%%%%%
\centerline {\bf REFERENCES}
%%%%%%%%%%%%%%%%%%%%%%%%%%%%%%%%%%%%%%%%%%%%%%%%%%%%%%%%%%%%%%%%%%


\begin{references}
\bibitem{protoncomment} For the proton $\sum_i e_i r^2_i=3/4m_q^2$, which is much larger than
the Foldy term and, by contributing $\simeq 0.27$ fm$^2$ to $r_{Ep}^2$, brings the constituent quark model prediction
for this  quantity into better agreement with
experiment \cite{HayneIsgur}.
\end{references}
\end{document}